 \documentclass[12pt]{article}         
\usepackage{graphics,amsmath,amssymb,feynmp}
\usepackage{thophys}
{\def\$#1: #2 ${#2}}

\begin{document}

\title{Towards a Complete Calculation of $\gamma\gamma \to 4f$.}

\author{%
  Edward Boos\thanks{%
      On leave from Moscow State University.}%
    \thanks{%
      Supported by Deutsche Forschungsgemeinschaft (DFG).}%
    \thanks{email:
      \texttt{boos@ifh.de}} \\
  Thorsten Ohl\thanks{%
       Supported by Bundesministerium f\"ur Bildung,
       Wissenschaft, Forschung und Technologie (BMBF), Germany.}%
    \thanks{e-mail:
      \texttt{Thorsten.Ohl@Physik.TH-Darmstadt.de}} \\
  \hfil \\
  Technische Hochschule Darmstadt \\
  Schlo\ss gartenstr. 9 \\
  D-64289 Darmstadt \\
  Germany}

\date{%
  IKDA 97/16\\
  hep-ph/9705374\\
  May 1997}

\maketitle
\begin{abstract}
  We propose a general classification of all four fermion final states
  in $\gamma\gamma$~collisions at a future $e^+e^-$~Linear
  Collider and discuss the relation of ``signal'' and ``background''
  contributions in vector boson production.  We utilize the results
  in a critical examination of the Higgs signal in sub-threshold $W^+W^-$
  pair production.
\end{abstract}

\begin{fmffile}{gg4fpics}

\section{Introduction}
\label{sec:introduction}

A $e^+e^-$ Linear Collider (LC) in the 500~GeV to 1~TeV range would
provide exciting physics opportunities~\cite{LCRev},
complementary to the Large
Hadron Collider (LHC) currently under construction.  In addition to
the $e^+e^-$ annihilation mode, such a collider would allow to study
physics in $e^-\gamma$ and $\gamma\gamma$ collisions as well.  The
photon beams can either be produced by Compton backscattering in a
dedicated photon collider~\cite{photon-collider},
but they are available, albeit with a softer spectrum, from
beamstrahlung~\cite{Chen/Noble:1986:Beamstrahlung} and bremsstrahlung
in any case.  Therefore numerically precise and theoretically reliable
calculations of $\gamma\gamma$ cross sections are required for
unleashing the physics potential of such a machine.

Standard model predictions for on-shell gauge boson production in
$\gamma\gamma$ collisions have been available for a long time in lowest
order~\cite{AA->WW}
and including electroweak radiative
corrections~\cite{Denner/Dittmaier/Schuster:1995:AA->WW/RC}.  However,
the observed final state in these processes is not the gauge boson
pair, but the four fermions that the gauge bosons decay into.  The
resonant diagrams that factorize into gauge boson production and decay
do \emph{not} form a gauge invariant subset.  Therefore a more
detailed investigation, including non-resonant ``background''
diagrams, is needed for obtaining theoretically consistent results.
Complete calculations for $e^+e^-\to4f$ have been performed in recent
years (see~\cite{e+e-->4f}
and references therein), but a similar
analysis of $\gamma\gamma\to 4f$ is not available yet.

For some gauge invariant subsets of the diagrams contributing to
$\gamma\gamma\to 4f$, first numerical results in the region above the
$W^+W^-$~threshold have been published
recently~\cite{CC13,CC21}.
These studies have focused on numerical calculations of cross sections
for specific final states and have not attempted a complete
investigation of $\gamma\gamma\to 4f$.  There have also been earlier
analytical calculations of the high energy asymptotics of
$\gamma\gamma\to\ell^+\ell^-e^+e^-$ in
QED~\cite{Kuraev/Schiller/Serbo:1985:gg_calibration}.
In order to pave the road for a
more systematic treatment, we will give a classification of all gauge
invariant subsets for $\gamma\gamma\to 4f$ in
section~\ref{sec:diagrams}.

\begin{figure}
  \begin{center}
    \hfill\\
    \vspace*{\baselineskip}
    \begin{fmfgraph*}(25,15)
      \fmfleft{g,g'}
      \fmflabel{$\gamma$}{g}
      \fmflabel{$\gamma$}{g'}
      \fmfright{w,w'}
      \fmflabel{$W^-$}{w}
      \fmflabel{$W^+$}{w'}
      \fmf{dbl_wiggly}{w',v',v,w}
      \fmf{photon}{g,v}
      \fmf{photon}{g',v'}
      \fmfdot{v,v'}
    \end{fmfgraph*}
    \qquad
    \begin{fmfgraph*}(25,15)
      \fmfleft{g,g'}
      \fmflabel{$\gamma$}{g}
      \fmflabel{$\gamma$}{g'}
      \fmfright{w,w'}
      \fmflabel{$W^-$}{w}
      \fmflabel{$W^+$}{w'}
      \fmf{dbl_wiggly}{w',v,w}
      \fmf{photon}{g,v}
      \fmf{photon}{g',v}
      \fmfdot{v}
    \end{fmfgraph*}
    \qquad
    \begin{fmfgraph*}(25,15)
      \fmfleft{g,g'}
      \fmflabel{$\gamma$}{g}
      \fmflabel{$\gamma$}{g'}
      \fmfright{w,w'}
      \fmflabel{$W^-$}{w}
      \fmflabel{$W^+$}{w'}
      \fmf{dbl_wiggly}{w',v',w}
      \fmf{photon}{g,v,g'}
      \fmf{dots,label=$H$,label.side=left}{v,v'}
      \fmfblob{4mm}{v}
      \fmfdot{v'}
    \end{fmfgraph*}
  \end{center}
  \caption{\label{fig:on-shell}%
    On-shell $W^+W^-$ production and Higgs signal.}
\end{figure}
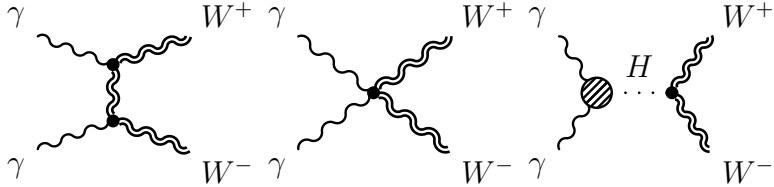

Recently, the search for the intermediate mass Higgs in the
reaction~$\gamma\gamma\to H\to W^+W^-$ below threshold, i.e.~with one
off-shell~$W$ has been discussed in~\cite{Ginzburg/Ivanov:1997:Higgs}.
This
scenario will provide us with an application of our classification.
It is obvious that in this range ($\sqrt s \approx M_H \lessapprox
2M_W$), the on-shell diagrams in figure~\ref{fig:on-shell} can serve
as a rough estimate only and that a more complete analysis is
required.  In section~\ref{sec:cross-sections} we will present
numerical results for this process and we will discuss their
phenomenological implications.

We will conclude with an outlook on the construction of a complete
Monte Carlo event generator for $\gamma\gamma\to 4f$, which is in
progress.

\section{Classification of Feynman Diagrams}
\label{sec:diagrams}

We start with a classification of all Feynman diagrams contributing to
four fermion production in $\gamma\gamma$~collisions.  This
classification has to be done from two perspectives: on one hand we
have to investigate the different topologies and singularity
structures for disentangling signals from backgrounds and for
constructing an efficient Monte Carlo event generator.  On the other
hand, we have to classify the flavor structure of the final states and
the gauge invariant subsets of diagrams corresponding to them.
In this classification we will use the analogue of the established
notation proposed for~$e^+e^-\to 4f$
in~\cite{Bardin/etal:1994:4f-classification}.

We will perform the classification for massless fermions and in
unitarity gauge first.  Later we will discuss the modifications
for massive fermions and for $R_\xi$~gauge. 

\subsection{Topologies and Singularities}
\label{sec:topologies}

\begin{figure}
  \begin{center}
    \hfill\\
    \vspace*{\baselineskip}
    \mbox{($Q$)
    \begin{fmfgraph*}(30,20)
      \fmfleft{g,g'}
      \fmflabel{$\gamma$}{g}
      \fmflabel{$\gamma$}{g'}
      \fmfright{f1,f2b,f3,f4b}
      \fmflabel{$f_1$}{f1}
      \fmflabel{$\bar f_2$}{f2b}
      \fmflabel{$f_3$}{f3}
      \fmflabel{$\bar f_4$}{f4b}
      \fmf{photon,t=2}{g,qgv,g'}
      \fmf{dbl_wiggly,lab=$W$,lab.side=left}{v1,qgv,v2}
      \fmf{fermion}{f2b,v1,f1}
      \fmf{fermion}{f4b,v2,f3}
      \fmfdot{v1,v2,qgv}
    \end{fmfgraph*}
    \qquad
    ($D$)
    \begin{fmfgraph*}(30,20)
      \fmfleft{g,g'}
      \fmflabel{$\gamma$}{g}
      \fmflabel{$\gamma$}{g'}
      \fmfright{f1,f2b,f3,f4b}
      \fmflabel{$f_1$}{f1}
      \fmflabel{$\bar f_2$}{f2b}
      \fmflabel{$f_3$}{f3}
      \fmflabel{$\bar f_4$}{f4b}
      \fmf{photon,t=2}{g,tgv1}
      \fmf{photon,t=2}{tgv2,g'}
      \fmf{dbl_wiggly,lab=$W$,lab.side=left}{v1,tgv1,tgv2,v2}
      \fmf{fermion}{f2b,v1,f1}
      \fmf{fermion}{f4b,v2,f3}
      \fmfdot{v1,v2,tgv1,tgv2}
    \end{fmfgraph*}
    \qquad
    ($T$)
    \begin{fmfgraph*}(30,20)
      \fmfleft{g,g'}
      \fmflabel{$\gamma$}{g}
      \fmflabel{$\gamma$}{g'}
      \fmfright{f1,f2b,f3,f4b}
      \fmflabel{$f_1$}{f1}
      \fmflabel{$\bar f_2$}{f2b}
      \fmflabel{$f_3$}{f3}
      \fmflabel{$\bar f_4$}{f4b}
      \fmf{photon,t=2}{g,v1}
      \fmf{photon,t=2}{tgv,g'}
      \fmffixedx{0}{v1,v2}
      \fmf{dbl_wiggly,lab=$W$,lab.side=left}{v2,tgv,v3}
      \fmf{fermion}{f2b,v2,v1,f1}
      \fmf{fermion}{f4b,v3,f3}
      \fmfdot{v1,v2,v3,tgv}
    \end{fmfgraph*}}\\[2\baselineskip]
    \mbox{($S$)
    \begin{fmfgraph*}(30,20)
      \fmfleft{g,g'}
      \fmflabel{$\gamma$}{g}
      \fmflabel{$\gamma$}{g'}
      \fmfright{f1,f4b,f3,f2b}
      \fmflabel{$f_1$}{f1}
      \fmflabel{$\bar f_2$}{f2b}
      \fmflabel{$f_3$}{f3}
      \fmflabel{$\bar f_4$}{f4b}
      \fmf{photon,t=2}{g,v1}
      \fmf{photon,t=2}{g',v3}
      \fmf{dbl_wiggly,lab=$W,,Z,,\gamma$}{v2,v4}
      \fmf{fermion}{f4b,v4,f3}
      \fmf{fermion}{f2b,v3,v2,v1,f1}
      \fmfdotn{v}{4}
    \end{fmfgraph*}
    \qquad
    ($B$)
    \begin{fmfgraph*}(30,20)
      \fmfleft{g,g'}
      \fmflabel{$\gamma$}{g}
      \fmflabel{$\gamma$}{g'}
      \fmfright{f1,f2b,f3,f4b}
      \fmflabel{$f_1$}{f1}
      \fmflabel{$\bar f_2$}{f2b}
      \fmflabel{$f_3$}{f3}
      \fmflabel{$\bar f_4$}{f4b}
      \fmf{photon,t=2}{g,v1}
      \fmf{photon,t=2}{g',v2}
      \fmffixedx{0}{v1,v2}
      \fmf{dbl_wiggly,lab=$W,,Z,,\gamma$,lab.side=left}{v3,v4}
      \fmf{fermion}{f2b,v3,v2,v1,f1}
      \fmf{fermion}{f4b,v4,f3}
      \fmfdotn{v}{4}
    \end{fmfgraph*}
    \qquad
    ($M$)
    \begin{fmfgraph*}(30,20)
      \fmfleft{g,g'}
      \fmflabel{$\gamma$}{g}
      \fmflabel{$\gamma$}{g'}
      \fmfright{f1,f2b,f3,f4b}
      \fmflabel{$f_1$}{f1}
      \fmflabel{$\bar f_2$}{f2b}
      \fmflabel{$f_3$}{f3}
      \fmflabel{$\bar f_4$}{f4b}
      \fmf{photon,t=2}{g,v1}
      \fmf{photon,t=2}{g',v4}
      \fmf{dbl_wiggly,lab=$W,,Z,,\gamma$,lab.side=left}{v2,v3}
      \fmf{fermion}{f2b,v2,v1,f1}
      \fmf{fermion}{f4b,v4,v3,f3}
      \fmfdotn{v}{4}
    \end{fmfgraph*}}
  \end{center}
  \caption{\label{fig:Q/D/T/S/B/M}%
    Topologies with a quartic gauge coupling ($Q$), with two triple
    gauge couplings ($D$), with one triple gauge coupling ($T$),
    with single vector boson production ($S$), with gauge
    boson bremsstrahlung ($B$) and multi-peripheral diagrams ($M$).}
\end{figure}
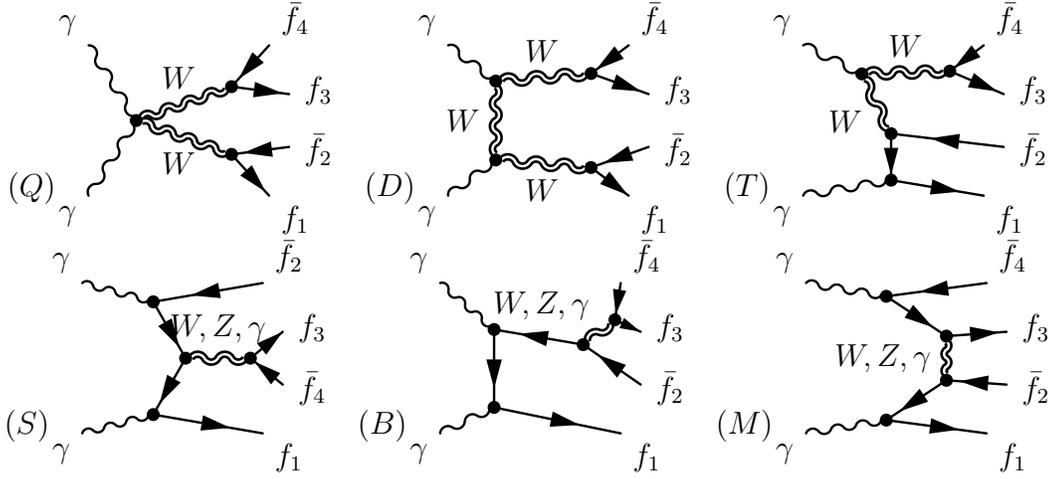

The Feynman diagrams for $\gamma\gamma\to4f$ have six different
topologies, which are depicted in figure~\ref{fig:Q/D/T/S/B/M}.
All diagrams can be derived from these six
topologies by charge conjugation and permutation of final state
fermions.

Each of these topologies corresponds to a particular singularity
structure and will be dominant in different regions of phase space:
$Q$ and $D$ have two resonant gauge boson propagators, while $T$, $S$
and $B$ have a single resonant gauge boson.  The multi-peripheral
contribution $M$ has no resonant gauge bosons at all.  An efficient
multi channel event generator will use these six topologies as
subchannels for mapping kinematical singularities.

\subsection{Final States and Gauge Invariant Subsets}
\label{sec:final-states}

While the unravelling of the singularity structures presented in the
previous section is essential for the construction of a Monte Carlo
event generator, the classification of the gauge invariant subsets is
more conveniently performed from a different point of view.

\begin{figure}
  \begin{center}
    \hfill\\
    \vspace*{\baselineskip}
    \begin{fmfgraph*}(30,15)
      \fmfbottom{d1,v1,d2,v2,d3}
      \fmftop{f1,f2b,f3,f4b}
      \fmflabel{$f_1$}{f1}
      \fmflabel{$\bar f_2$}{f2b}
      \fmflabel{$f_3$}{f3}
      \fmflabel{$\bar f_4$}{f4b}
      \fmf{dbl_wiggly,lab=$W,,Z,,\gamma$,lab.side=right}{v1,v2}
      \fmf{fermion}{f2b,v1,f1}
      \fmf{fermion}{f4b,v2,f3}
      \fmfdot{v1,v2}
    \end{fmfgraph*}
  \end{center}
  \caption{\label{fig:0to4f}%
    Amplitude for $\ket{0} \to f_1 \bar f_2 f_3 \bar f_4$, from which
    all diagrams are derived.}
\end{figure}
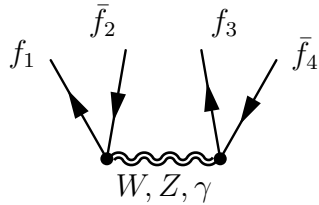

Since the photon couplings are diagonal in flavor space, the flavor
structure of the four fermion final states can be investigated easily
by looking at the amplitude for~$\ket{0} \to f_1 \bar f_2 f_3 \bar
f_4$ (see figure~\ref{fig:0to4f}).  The gauge invariant\footnote{%
   Gauge invariant for vanishing gauge boson width.  We are not yet
   concerned with the intricate interplay of resummation and gauge
   invariance~\cite{BHF}.}
subsets of diagrams are then obtained by hooking two photons to
the charged propagators in all possible ways.

For two generations and a diagonal CKM matrix, there are alltogether
45~different four fermion final states in $\gamma\gamma$~collisions.
Twelve of these are six pairs of charge conjugate states.  The third
generation will add 53~final states, which are trivial replicas of
final states appearing in the two generation case.  Therefore we will
restrict our discussion to the case of two generations.

\begin{table}
  \begin{center}
    \begin{tabular}{|c|c|cccc|c|}\hline
      \multicolumn{2}{|c|}{Class}
                    & \multicolumn{4}{|c|}{Final States}
                    & $3^{\text{rd}}$ Gen.\\\hline\hline
      \textit{CC13} & l  & $e^- \bar\nu_{e} \mu^+ \nu_\mu     $
                         & $e^+ \nu_{e}     \mu^- \bar\nu_\mu $
                         &
                         &
                         & $+4$ \\\hline
      \textit{CC21} & sl & $e^- \bar\nu_{e} u \bar d $
                         & $e^+ \nu_{e} \bar u d $
                         & $e^- \bar\nu_{e} c \bar s $
                         & $e^+ \nu_{e} \bar c s $
                         & \\\hline
      \multicolumn{2}{|c|}{}
                         & $\mu ^- \bar\nu_{\mu } u \bar d $
                         & $\mu ^+ \nu_{\mu } \bar u d $
                         & $\mu ^- \bar\nu_{\mu } c \bar s $
                         & $\mu ^+ \nu_{\mu } \bar c s $
                         & $+10$ \\\hline
      \textit{CC31} & h  & $u \bar d \bar c s $
                         & $d \bar u c \bar s $
                         &
                         &
                         & $+4$ \\\hline
    \end{tabular}
  \end{center}
  \caption{\label{tab:CC}%
    The 12~\textit{CC$n$} final states.  Including a third generation
    increases the count to~30.  In this and the following tables, we
    tag leptonic, semileptonic and hadronic final states with~`l',
    `sl' and~`h' respectively.}
\end{table}

\begin{table}
  \begin{center}
    \begin{tabular}{|c|c|cccc|c|}\hline
      \multicolumn{2}{|c|}{Class}
                    & \multicolumn{4}{|c|}{Final States}
                    & $3^{\text{rd}}$ Gen.\\\hline\hline
      \textit{NC06} & l  & $\nu_{e} \bar\nu_{e} \mu ^- \mu ^+ $
                         & $\nu_{\mu } \bar\nu_{\mu } e^- e^+ $
                         &
                         &
                         & $+4$ \\\hline
      \textit{NC06} & sl & $\nu_{e} \bar\nu_{e} u \bar u $
                         & $\nu_{e} \bar\nu_{e} d \bar d $
                         & $\nu_{e} \bar\nu_{e} c \bar c $
                         & $\nu_{e} \bar\nu_{e} s \bar s $
                         & \\\hline
      \multicolumn{2}{|c|}{}
                         & $\nu_{\mu } \bar\nu_{\mu } u \bar u $
                         & $\nu_{\mu } \bar\nu_{\mu } d \bar d $
                         & $\nu_{\mu } \bar\nu_{\mu } c \bar c $
                         & $\nu_{\mu } \bar\nu_{\mu } s \bar s $
                         & $+10$ \\\hline
      \textit{NC20} & l  & $e^- e^+ e^- e^+ $
                         & $\mu ^- \mu ^+ \mu ^- \mu ^+ $
                         &
                         &
                         & $+1$ \\\hline
      \textit{NC20} & h  & $u \bar u u \bar u $
                         & $d \bar d d \bar d $
                         & $c \bar c c \bar c $
                         & $s \bar s s \bar s $
                         & $+2$ \\\hline
      \textit{NC40} & l  & $e^- e^+ \mu ^- \mu ^+ $
                         &
                         &
                         &
                         & $+2$\\\hline
      \textit{NC40} & sl & $e^- e^+ u \bar u $
                         & $e^- e^+ d \bar d $
                         & $e^- e^+ c \bar c $
                         & $e^- e^+ s \bar s $
                         & \\\hline
      \multicolumn{2}{|c|}{}
                         & $\mu ^- \mu ^+ u \bar u $
                         & $\mu ^- \mu ^+ d \bar d $
                         & $\mu ^- \mu ^+ c \bar c $
                         & $\mu ^- \mu ^+ s \bar s $
                         & $+10$ \\\hline
      \textit{NC40} & h  & $u \bar u s \bar s $
                         & $u \bar u c \bar c $
                         & $d \bar d s \bar s $
                         & $d \bar d c \bar c $
                         & $+4$ \\\hline
    \end{tabular}
  \end{center}
  \caption{\label{tab:NC}%
    The 29~\textit{NC$n$} final states.  Including a third generation
    increases the count to~62.}
\end{table}

\begin{table}
  \begin{center}
    \begin{tabular}{|c|c|cc|c|}\hline
      \multicolumn{2}{|c|}{Class}
                     & \multicolumn{2}{|c|}{Final States}
                     & $3^{\text{rd}}$ Gen.\\\hline\hline
      \textit{mix19} & l  & $e^- e^+ \nu_{e} \bar\nu_{e} $
                          & $\mu ^- \mu ^+ \nu_{\mu } \bar\nu_{\mu } $
                          & $+1$ \\\hline
      \textit{mix71} & h  & $u \bar u d \bar d $
                          & $c \bar c s \bar s $
                          & $+1$ \\\hline
    \end{tabular}
  \end{center}
  \caption{\label{tab:mix}%
    The 4~\textit{mix$n$} final states.  Including a third generation
    increases the count to~6.}
\end{table}

Obviously, the sets $f_1 \bar f_2 f_3 \bar f_4$ fall into three
classes.  The first class contains 12 final states (\textit{CC$n$}, see
table~\ref{tab:CC}) where the two pairs exchange a~$W$.  They consist
of two pairs of particles with the anti-particle of their partner from
the weak isospin dublett.  The second class contains 29 final states
(\textit{NC$n$}, see table~\ref{tab:NC}) that exchange a neutral gauge boson,
$Z$ or $\gamma$.  They consist of two pairs of particles with their
antiparticles.  Finally there are four final states (\textit{mix$n$}, see
table~\ref{tab:mix}) that can exchange both charged and neutral gauge
bosons.

Counting the number of diagrams for processes in each class is simple.
Obviously, it can only depend on the number~$n_c$ of charged fermions.
For the \textit{CC$n$} case, there are $n_c+1$ charged propagators to
attach the
first photon to and consequently $n_c+2$ for the second photon.
Adding the single diagram with a $W^+W^-\gamma\gamma$~vertex, we find
\begin{equation}
  N_{CC} = n_c^2 + 3n_c + 3  \qquad\qquad (n_c>0) \,.
\end{equation}
{}From charge conservation, only the cases~$n_c=2,3,4$ are realized,
leading to the sets \textit{CC13}, \textit{CC21} and \textit{CC31},
respectively.

In the \textit{NC$n$} case, there are $n_c$ charged propagators for the first
photon and $n_c+1$ for the second.  Taking into account that both $Z$
and $\gamma$ can be exchanged for $n_c=4$, we can write
\begin{equation}
  N_{NC} = \frac{n_c^2(n_c+1)}{2} \qquad\qquad (n_c\in\{0,2,4\})\,.
\end{equation}
Only the cases~$n_c=2,4$ are realized, leading to the sets \textit{NC06}
and \textit{NC40}.  For final states involving identical particles,
the \textit{NC40} degenerates to a \textit{NC20}.

Finally, the \textit{CC13} and \textit{NC06} can be combined to
\textit{mix19} and \textit{CC31} and \textit{NC40} can be combined to
\textit{mix71}.

\begin{table}
  \begin{center}
    \begin{tabular}{|c||c|c|c|}\hline
      $n_c$ & \textit{CC}   & \textit{NC}   & \textit{mix}   \\\hline\hline
      0     & ---           & ---           & ---            \\\hline
      1     & ---           & ---           & ---            \\\hline 
      2     & \textit{CC13} & \textit{NC06} & \textit{mix19} \\\hline
      3     & \textit{CC21} & ---           & ---            \\\hline
      4     & \textit{CC31} & \textit{NC40} & \textit{mix71} \\\hline
      4'    & ---           & \textit{NC20} & ---            \\\hline
    \end{tabular}
  \end{center}
  \caption{\label{tab:counting}%
    Counting diagrams.}
\end{table}

It is now a straightforward combinatorial exercise to determine the
number of Feynman diagrams from each topology contributing for a
specific final state.  The result is shown in table~\ref{tab:classes}.

\begin{table}
  \begin{center}
    \begin{tabular}{|c|c|r|r|r|r|r|r|}\hline
      \multicolumn{2}{|c|}{Class}
                              & Q &  T &  D &  S &  B &  M \\\hline\hline
      \textit{CC13}  & l      & 1 &  4 &  2 &  0 &  4 &  2 \\\hline
      \textit{CC21}  & sl     & 1 &  6 &  2 &  2 &  6 &  4 \\\hline
      \textit{CC31}  & h      & 1 &  8 &  2 &  4 &  8 &  8 \\\hline
      \textit{NC06}  & l/sl   & 0 &  0 &  0 &  2 &  4 &  0 \\\hline
      \textit{NC20}  & l/h    & 0 &  0 &  0 &  4 &  8 &  8 \\\hline
      \textit{NC40}  & l/sl/h & 0 &  0 &  0 &  8 & 16 & 16 \\\hline
      \textit{mix19} & l      & 1 &  4 &  2 &  2 &  8 &  2 \\\hline
      \textit{mix71} & h      & 1 &  8 &  2 & 12 & 24 & 24 \\\hline
    \end{tabular}
  \end{center}
  \caption{\label{tab:classes}%
    The eight classes of diagrams in $\gamma\gamma\to 4f$ and the
    corresponding topologies.  Note that $\text{\textit{NC40}} =
    2\cdot\text{\textit{NC20}}$, $\text{\textit{mix19}} =
    \text{\textit{NC06}} + \text{\textit{CC13}}$ and 
    $\text{\textit{mix71}} = \text{\textit{NC40}} + \text{\textit{CC31}}$.}
\end{table}
\subsection{Massive Fermions}
\label{sec:mass}

The modifications for massive fermions in unitarity gauge are
trivial: iff all four fermions are massive, the sets \textit{NC20}
and~\textit{NC40} are augmented by Higgs exchange diagrams to
\textit{NC30} and~\textit{NC60}.  All final states corresponding to
the set \textit{NC06} contain neutrinos (see table~\ref{tab:NC}) and
remain unaffected.  The charged current final states remain unaffected
by the Higgs at tree level as well.

\subsection{$R_\xi$~Gauge}
\label{sec:Rxi}

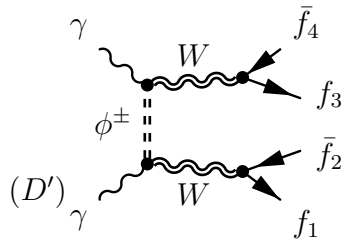
\begin{figure}
  \begin{center}
    \hfill\\
    \vspace*{\baselineskip}
    ($D'$)
    \begin{fmfgraph*}(30,20)
      \fmfleft{g,g'}
      \fmflabel{$\gamma$}{g}
      \fmflabel{$\gamma$}{g'}
      \fmfright{f1,f2b,f3,f4b}
      \fmflabel{$f_1$}{f1}
      \fmflabel{$\bar f_2$}{f2b}
      \fmflabel{$f_3$}{f3}
      \fmflabel{$\bar f_4$}{f4b}
      \fmf{photon,t=2}{g,tgv1}
      \fmf{photon,t=2}{tgv2,g'}
      \fmf{dbl_dashes,lab=$\phi^\pm$,lab.side=left}{tgv1,tgv2}
      \fmf{dbl_wiggly,lab=$W$,lab.side=left}{v1,tgv1}
      \fmf{dbl_wiggly,lab=$W$,lab.side=left}{tgv2,v2}
      \fmf{fermion}{f2b,v1,f1}
      \fmf{fermion}{f4b,v2,f3}
      \fmfdot{v1,v2,tgv1,tgv2}
    \end{fmfgraph*}
  \end{center}
  \caption{\label{fig:D'}%
    Sole additional diagram in $R_\xi$~gauge for massless fermions.}
\end{figure}

In a $R_\xi$~gauge, additional diagrams involving the charged and
neutral Goldstone bosons~$\phi^\pm,\phi^0$ have to be taken into
account.  Since the couplings of the Goldstone bosons are proportional
to the masses of the participating particles, only the diagram~$D'$
depicted in figure~\ref{fig:D'} has to be added for massless fermions in
the final state.

\begin{figure}
  \begin{center}
    \hfill\\
    \vspace*{\baselineskip}
    ($\fmslash{Q}$)
    \begin{fmfgraph*}(30,20)
      \fmfleft{g,g'}
      \fmflabel{$\gamma$}{g}
      \fmflabel{$\gamma$}{g'}
      \fmfright{f1,f2b,f3,f4b}
      \fmflabel{$f_1$}{f1}
      \fmflabel{$\bar f_2$}{f2b}
      \fmflabel{$f_3$}{f3}
      \fmflabel{$\bar f_4$}{f4b}
      \fmf{photon,t=2}{g,qgv,g'}
      \fmf{dbl_wiggly,lab=$W^\mp$,lab.side=left}{v1,qgv}
      \fmf{dbl_dashes,lab=$\phi^\pm$,lab.side=left}{qgv,v2}
      \fmf{fermion}{f2b,v1,f1}
      \fmf{fermion}{f4b,v2,f3}
      \fmfdot{v1,v2,qgv}
    \end{fmfgraph*}
  \end{center}
  \caption{\label{fig:AAWphi-absent}%
    Exception to the rule that gauge bosons can be replaced by their
    corresponding Goldstone bosons individually.  This diagram and its
    charge conjugate are \emph{absent}.} 
\end{figure}
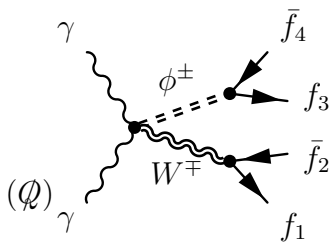

For massive fermions, the situation is more involved, but it can be
summarized in the rule that each massive gauge boson that is connected
to massive particles at \emph{both} ends can be replaced by its
corresponding Goldstone boson individually.  The only exception to
this rule is that there is no $W^\mp\phi^\pm\gamma\gamma$~vertex.
Therefore the diagram~$\fmslash{Q}$ depicted in
figure~\ref{fig:AAWphi-absent} and its charge conjugate are
\emph{absent}.

\section{Cross Sections}
\label{sec:cross-sections}

For a numerical illustration of the importance of ``background''
diagrams, we present results for cross sections for two \textit{CC$n$}
final states.  The calculations have been performed by means of the
\texttt{CompHEP} system~\cite{CompHEP}.
We present the results as a function of
the $\gamma\gamma$~invariant mass~$\sqrt{s'}$.

\subsection{CC13: $e^+ \mu ^- \bar\nu_{\mu } \nu_{e} $}
\label{sec:CC13}

This state has a very clean signature: a muon, a positron, missing
energy and no other activity in the detector.  It is the smallest
gauge invariant subset containing resonant $W$~pair production.
Nevertheless, it involves all topologies except one.  Cross sections
for this final state have been calculated earlier
in~\cite{CC13}.

\begin{figure}
  \begin{center}
    \includegraphics{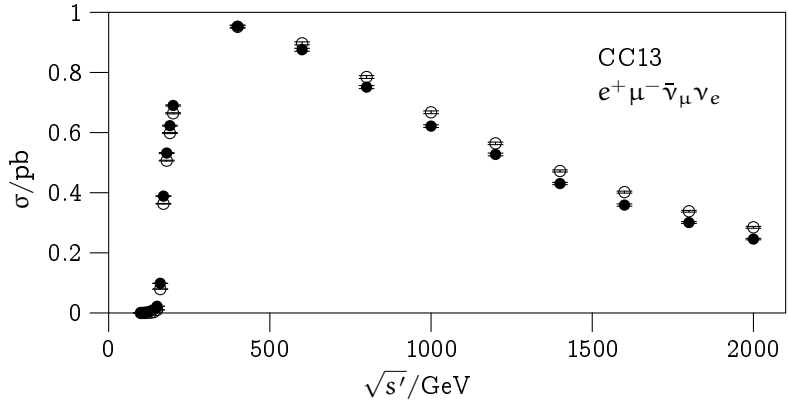}
  \end{center}
  \caption{\label{fig:xsect-cc13}%
    \textit{CC13} cross section (full circles) and signal
    contribution (open circles), after the canonical LC cuts:
    $E_{\text{lepton}} > 1\,\text{GeV}$,
    $\theta_{\text{lepton}} > 10\,\text{deg}$,
    $E_{\text{jet}} > 3\,\text{GeV}$,
    $\theta_{\text{jet}} > 5\,\text{deg}$,
    $\theta_{\text{lepton},\text{lepton}'} > 5\,\text{deg}$,
    $\theta_{\text{lepton},\text{jet}} > 5\,\text{deg}$ and
    $m_{\text{jet},\text{jet}'} > 10\,\text{GeV}$.
    The 1\,sd error bars are shown, unless they are smaller than the
    thickness of the lines.}
\end{figure}

Figure~\ref{fig:xsect-cc13} shows the total cross section for
\textit{CC13} with the canonical cuts for the DESY/ECFA Linear
Collider study (our results agree
with~\cite{CC13}
in their cuts).  Since this final state includes two neutrinos, no 
invariant mass cut can be applied to suppress the ``background''.
Clearly, the cross section is dominated above threshold by the
``signal''.  However, an excess in this channel would be a signal for
rather exotic flavor changing ``new physics''.  Therefore, the
standard model contribution has to be known to high accuracy,
including ``background'' contributions.

\subsection{CC21: $e^- \bar\nu_{e} u \bar d $}
\label{sec:CC21}

\begin{figure}
  \begin{center}
    \includegraphics{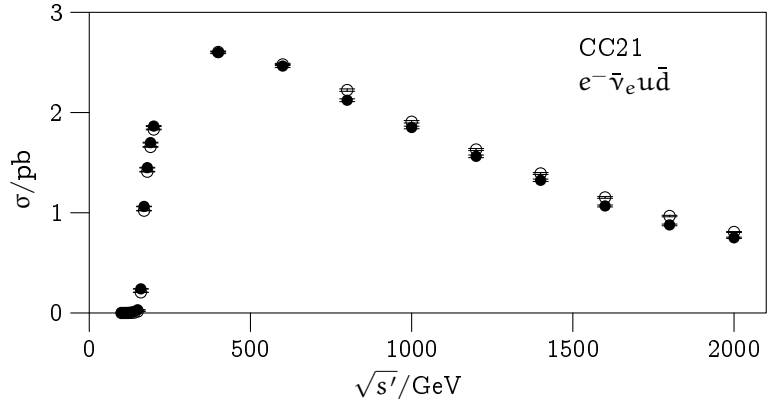}
  \end{center}
  \caption{\label{fig:xsect-cc21}%
    \textit{CC21} cross section (full circles) and signal
    contribution (open circles), after the canonical NLC cuts.
    An invariant mass cut
    of~$M_W-10\text{GeV}\le\sqrt{(p_1+p_2)^2}\le M_W+10\text{GeV}$
    has been applied to both cross sections.}
\end{figure}

The \textit{CC21} diagrams have been calculated before
in~\cite{CC21}.
Figure~\ref{fig:xsect-cc21} shows the total cross section for
\textit{CC21} with the canonical cuts for the DESY/ECFA Linear
Collider study (using the cuts of~\cite{CC21}
we find agreement again).
Here we can also apply an invariant mass cut for the
two jets. Such a cut will be used in the experiment for event
selection and it will suppress background diagrams.  In
figure~\ref{fig:xsect-cc21} we have applied the
cut~$M_W-10\text{GeV}\le\sqrt{(p_1+p_2)^2}\le M_W+10\text{GeV}$ to
both signal and full cross section.
Again, the cross section is dominated above threshold by the
``signal'', but for detailed studies of the vector boson couplings,
the ``background'' contributions cannot be neglected.

\subsection{Sub-Threshold Cross sections}
\label{sec:sub-thresh}

\begin{figure}
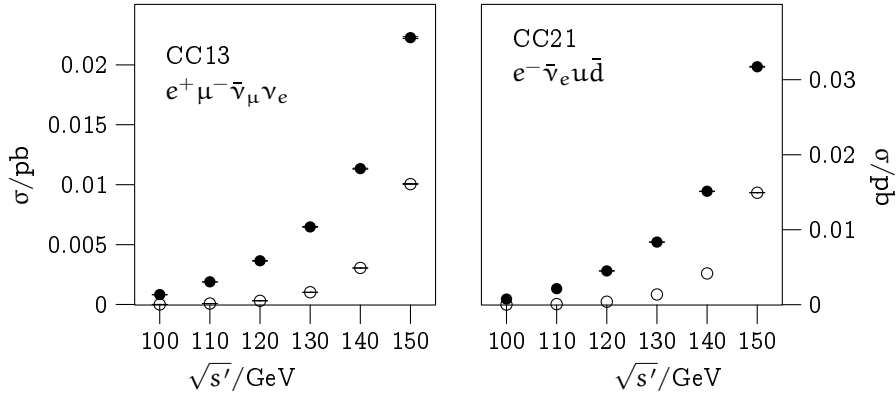

  \begin{center}
    \includegraphics{plots.3}\quad
    \includegraphics{plots.4}
  \end{center}
  \caption{\label{fig:xsect-threshold}%
    Sub-threshold cross section (full circles) and signal
    contribution (open circles).}
\end{figure}

In figure~\ref{fig:xsect-threshold} we have enlarged the region below
the $W^+W^-$ threshold.  In this region, which can be important in the
search for an light intermediate mass Higgs, the cross section is
several times larger than the prediction from the ``signal diagrams''
only.  Therefore, the signal to background ratios predicted in studies
based on ``signal diagrams'' only
(cf.~\cite{Ginzburg/Ivanov:1997:Higgs}) have to be reduced
considerably.  Nevertheless, since the standard model contributions
can be calculated completely, this is only a technical obstacle
pointing to the need for complete calculations.

\section{Conclusions}
\label{sec:conclusions}

We have proposed a complete classification of four fermion final
states in $\gamma\gamma$~collisions.  As an application, we have
demonstrated the need for full, gauge invariant calculations for
obtaining reliable predictions below the $W^+W^-$ threshold.  The
restriction of calculations to the ``signal diagrams'' is not
sufficient. Therefore the signal to background ratios presented
in~\cite{Ginzburg/Ivanov:1997:Higgs} turn out to be far too
optimistic.  Nevertheless, once a complete calculation is available,
the potential of the Linear Collider for the intermediate mass Higgs
can be realized.

While the results presented in this note correspond to full
calculations and are gauge invariant, they can only be a first step. A
Monte Carlo event generator will be required for more detailed
experimental studies.  Only such a tool will allow the efficient
simulation of~$\gamma\gamma\to 4f$ for arbitrary experimental cuts.

The helicity amplitudes and the phase space for massless fermions are
relatively simple.  Therefore, the construction of a Monte Carlo event
generator for~$\gamma\gamma\to 4f$ with massless fermions will be the
natural next step~\cite{Boos/Ohl:1997:MC} towards a complete calculation.


\end{fmffile}
\end{document}